 \documentclass[structabstract]{aa}
\usepackage{txfonts}
\usepackage{natbib}
\bibpunct{(}{)}{,}{a}{}{,} 
\usepackage{graphicx}

\begin{document}
\title{Modeling the ionizing spectra of  \ion{H}{ii} regions: individual stars versus stellar ensembles}
\author{Marcos Villaverde\inst{1}  \and Miguel Cervi\~no\inst{1} \and Valentina Luridiana\inst{2,3,1}}
\institute{Instituto de Astrof{\'\i}sica de Andaluc{\'\i}a (CSIC). Camino bajo de Hu{\'{e}}tor, 50, Granada 18080, Spain \and Instituto de Astrof{\'\i}sica de Canarias,
C/ V{\'\i}a L\'actea, s/n, 38205 La Laguna, Spain \and Departamento de Astrof{\'\i}sica, Universidad de La Laguna (ULL), E-38205 La Laguna, Tenerife, Spain}
\date{ Received 27 July 2009 / Accepted 4 June 2010}
\abstract {}
{We study how IMF sampling affects the ionizing flux and emission line spectra of low mass stellar clusters.}
{We performed $2\times 10^6$ Monte Carlo simulations of zero-age solar-metallicity stellar clusters covering the $20 - 10^6$ M$_{\sun}$ mass range. We study the distribution of cluster stellar masses, $M_\mathrm{clus}$, ionizing fluxes,  $Q(\mathrm{H}^0)$, and effective temperatures, $T_\mathrm{eff}^\mathrm{clus}$. 
We compute photoionization models that broadly describe the results of the simulations and compare them with photoionization grids.}
{Our main results are: (a) A large number of low mass clusters (80\% for $M_\mathrm{clus}$ = 100 M$_{\sun}$) are unable to form an H~{\sc ii} region. (b) There are a few overluminous stellar clusters that form H~{\sc ii} regions. These overluminous clusters preserve statistically the mean value of $\langle Q(\mathrm{H}^0)\rangle$ obtained by synthesis models, but the mean value cannot be used as a description of particular clusters. (c) The ionizing continuum of clusters with  $M_\mathrm{clus} \lesssim 10^4$ M$_{\sun}$ is more accurately described by an $individual$ star with self-consistent effective temperature ($T_\mathrm{eff}^\mathrm{*}$) and $Q(\mathrm{H}^0)$ than by the ensemble of stars (or a cluster $T_\mathrm{eff}^\mathrm{clus}$) produced by synthesis models. (d) Photoionization grids of stellar clusters can not be used to derive the global properties of low mass clusters.}
{Although  variations in the upper mass limit,  $m_\mathrm{up}$, 
of the IMF would reproduce the effects of IMF sampling, we find that an $ad ~hoc$ law that relates $m_\mathrm{up}$ to $M_\mathrm{clus}$ in the modelling of stellar clusters is useless, since: (a) it does not cover the whole range of possible cases, and (b) the modelling of stellar clusters with an IMF is motivated by the need to derive the $global$ properties of the cluster: however, in clusters affected by sampling effects we have no access to global information of the cluster but only particular information about a few individual stars.}
\authorrunning{M. Villaverde, M. Cervi\~no \& V. Luridiana}
\titlerunning{Individual stars vs. stellar ensembles}

\maketitle

\section{Introduction}

\begin{figure}[ht]
\includegraphics[width=8.5cm]{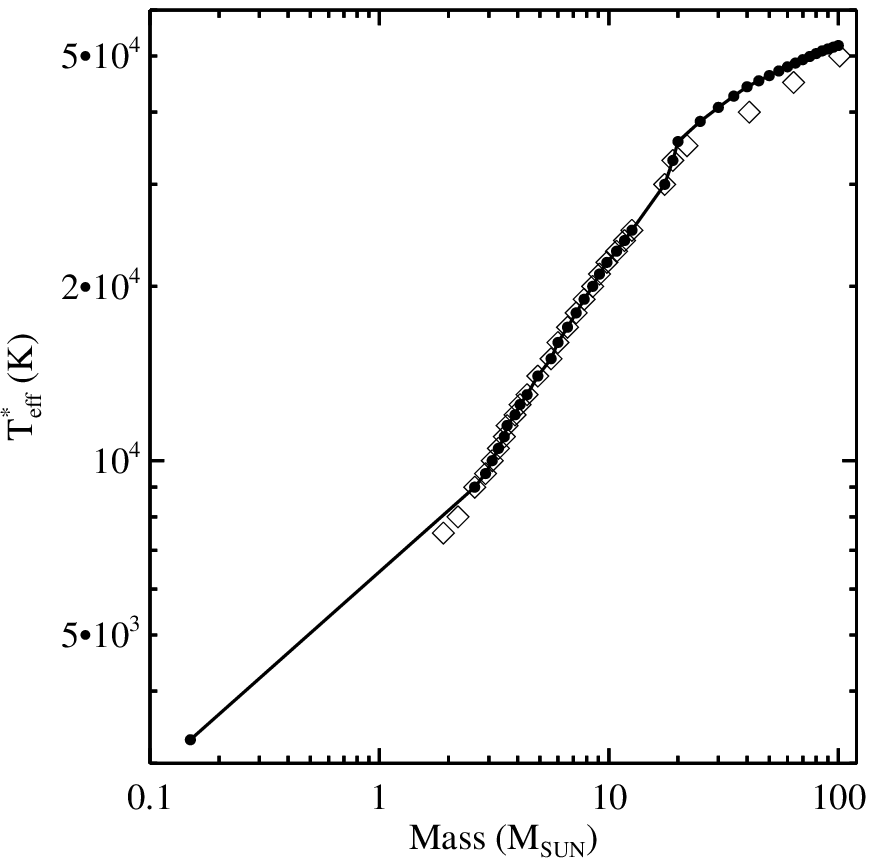}
\caption{
Connected dots: $T_{\mathrm{eff}}^\mathrm{*}$ vresus (vs.) stellar mass ($m$) relation for individual stars assumed in this work. Open diamonds: $T_{\mathrm{eff}}^\mathrm{*}$ vs. stellar mass ($m$) relation from \citet{DMetal98} data.}
\label{fig:m-T}
\end{figure}
\begin{figure}[ht]
\includegraphics[width=8.5cm]{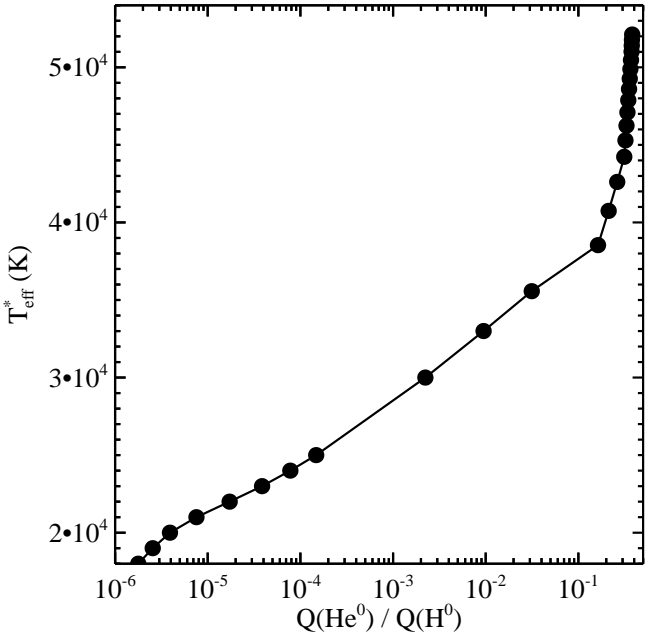}
\caption{$T_{\mathrm{eff}}^\mathrm{*}$ vs. $Q(\mathrm{He}^0)/Q(\mathrm{H}^0)$ ratio for individual stars assumed in this work.}
\label{fig:Qratio-T}
\end{figure}

\ion{H}{ii} regions are gas clouds photoionized by the ionizing radiation of close-by stars and they are identified observationally with their characteristic emission-line spectrum. This spectrum depends on the physical conditions of the gas and the shape and intensity of the stellar radiation field, which depend in turn on the evolutionary status of the stellar component. However, in the general astrophysical context, we have access neither to the evolutionary status nor the physical conditions of the components, and we must derive them from observations. However, the overall problem is difficult (and sometime impossible) to solve in an exact way because of the large number of unknowns involved, such as the exact 3-D gas and stellar distributions, the interrelations and feedback between the different components, and in general, the unknown  prior evolution of the system.  

One way to determine the physical conditions of the system is to perform a tailored analysis of the observed object \cite[e.g.,][]{Luretal99,Luretal01,Luretal03}. For large samples of data, it would be more economic (in terms of time and personal effort) to assume a set of hypotheses that simplify the problem as much as possible but which, at the same time, allow us to obtain results as realistic and generic as possible. 

In terms of the physical conditions of the gas, a common simplifying hypothesis is to assume a given photoionization case, such as cases A, B, C, or D \citep{Luretal09} for the observed object or set of objects. For the evolutionary status of the stellar component, there are common simplifying hypotheses for the description of the ionizing continuum produced by the cluster. Examples are methods using either individual stars, such as the use of a cluster effective temperature \cite[e.g.,][]{Sta90, SS97, BKG99, BK02} or a stellar ensemble where the different stellar components are mixed according to some specific prescriptions, as in evolutionary synthesis models (\citealt{Olof89,GVD94} are examples of pioneering works and \citealt{Detal06} a more recent one).

Although  the ionizing continuum cluster may be more accurately represented by the ensemble of stars deterministically predicted by a sophisticated synthesis model, this may not necessarily be the correct approach. Evolutionary synthesis models assume average numbers of stars with different effective temperatures and luminosities determined by the stellar evolution and the stellar birth rate: the number of stars with a given mass that were born at a given time (parametrized by the initial mass function, IMF, and the star formation history, SFH). The averages correctly describe the asymptotic properties of  clusters with infinite (or a very large) number of stars \cite[see ][ for more detailed explanations]{CVG09,CL05,CL06}. In more realistic cases, this proportionality is only valid as the average of several clusters, but it is not always a reliable representation of {\it individual clusters}.
In the case of low mass clusters ($M_\mathrm{clus} \lesssim 10^3$ M$_{\sun}$), the number of stars is also not large enough to properly sample the IMF, which may produce biased results when the observations are analysed by comparing them to the mean value obtained by synthesis models \citep{CVG03}. These effects are expected to exist in a great part of the ionizing clusters, since the majority of the embedded clusters have low masses \cite[M$_\mathrm{clus} < 10^3 $M$_{\sun}$,][]{LL03}. 

Given the decay of the IMF at high masses, the net effect is that a large number of low mass clusters ({\it but not all}) will not produce massive stars. In these cases, we demostrate that the spectrum of an individual star can provide a closer fit to the cluster ionizing continuum than the result of a synthesis model used in a deterministic way.

In this paper, we show, by means of Monte Carlo simulations of {\it zero age} stellar clusters, that between these two approaches (individual star versus (vs.) ensemble representations) there is a smooth transition that depends on the cluster mass. We establish how representative individual stellar spectra are of low cluster masses and show their implications for estimating the physical properties of clusters by means of their emission line spectra.

In section 2, we present our Monte Carlo simulations and probability computations. In section 3, we describe the photoionization simulations and analyse the results by means of diagnostic diagrams. The  implications of the analysis of the results is discussed in section 4. Finally, in section 5 we summarize the main conclusions of this work.

\section{The distribution of the ionizing properties of clusters}
\label{sec:ionMontecarlo}

\begin{figure*}[ht]
\includegraphics[width=\textwidth]{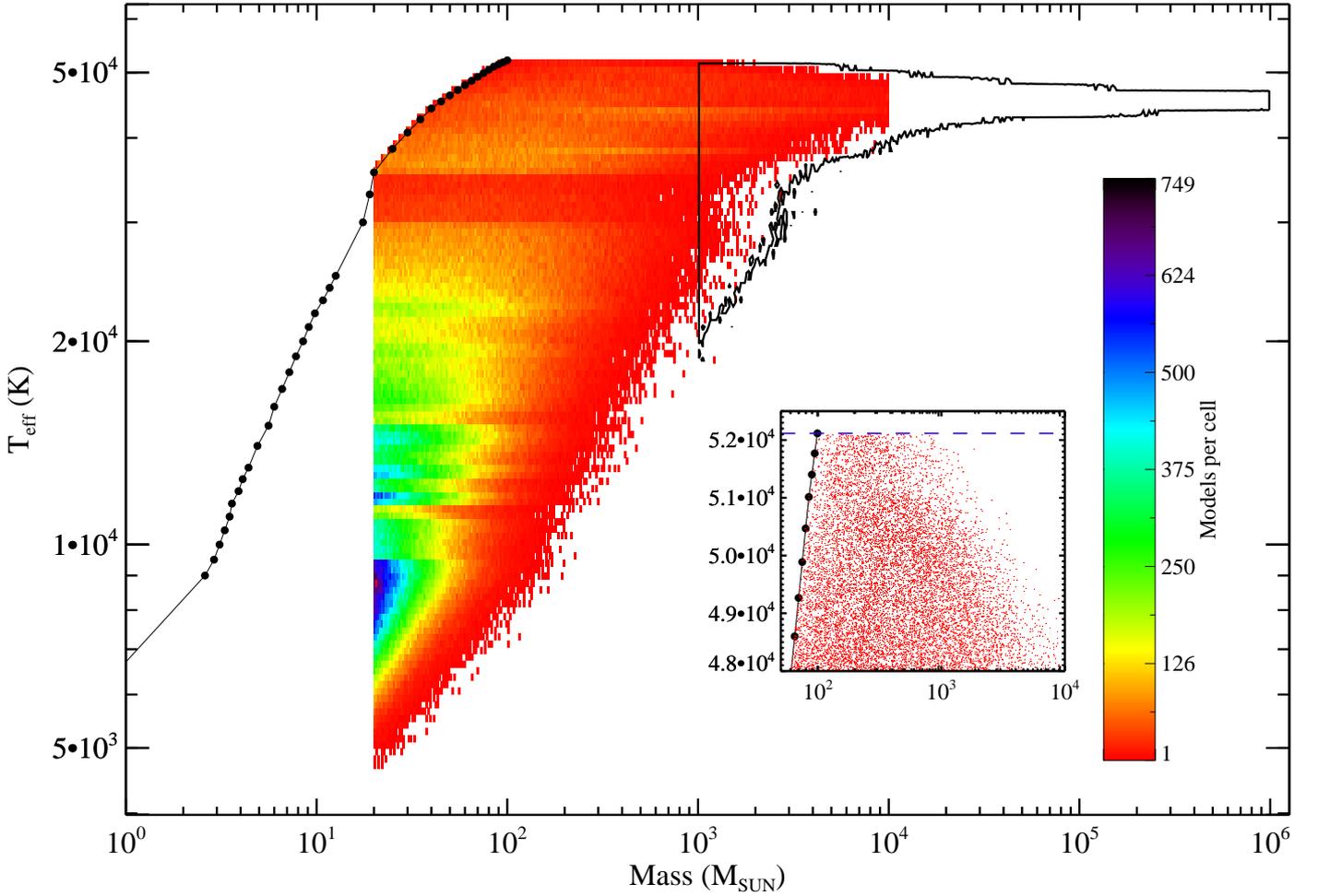}
\caption{The color-coded area shows the $M_\mathrm{clus} - T_{\mathrm{eff}}^\mathrm{clus}$ distribution for the low-mass cluster set. The solid line with dots on the left of the plot is the $m - T_{\mathrm{eff}}^\mathrm{*,ref}$ reference scale plotted in Fig. \ref{fig:m-T}. Note that the stellar $m - T_{\mathrm{eff}}^\mathrm{*,ref}$ reference scale defines the low mass limit of the $M_\mathrm{clus} - T_{\mathrm{eff}}^\mathrm{clus}$ distribution. The solid-line contour on the right of the plot, reaching  $M_\mathrm{clus}$ higher than $10^4$ M$_{\sun}$, corresponds to the simulations of the high-mass cluster set (c.f. Fig. \ref{fig:Mcl-Teff_high}). The inner box shows a detail of the region around $100$ M$_{\sun}$, where the maximum value of the cluster $T_{\mathrm{eff}}^\mathrm{clus}$ is reached.}
\label{fig:Mcl-Teff_low}
\end{figure*}

The first step in characterizing the emission line spectra of H~{\sc ii} regions is to attempt to reproduce their ionizing continuum. For this purpose, we used the cluster effective temperature, $T_{\mathrm{eff}}^\mathrm{clus}$, instead of a detailed spectral energy distribution.  The cluster $T_{\mathrm{eff}}^\mathrm{clus}$ is defined as
the effective temperature of a {\it reference} star ($T_{\mathrm{eff}}^\mathrm{*,ref}$) that has
the same ratio $Q(\mathrm{He}^{0})/Q(\mathrm{H}^{0})$ as the cluster \citep{M-HK91}, i.e., 
the same number of ionizing photons emitted below the {He\sc{ i}} Lyman limit at 504{~\AA}, $Q(\mathrm{He}^{0})$, with respect to the total ionizing photon number, $Q(\mathrm{H}^{0})$.

The reference scale was defined in the following way:

\begin{enumerate}
\item The relation between $Q(\mathrm{H}^0)$ and initial mass was obtained using the ZAMS values published by  \citet{DMetal98} (their Table 1), after excluding their two lowest mass points (see below).
 \item Low mass stars ($m \leq 20 {\mathrm{M}}_{\sun}$):
\begin{enumerate}
 \item $T_{\mathrm{eff}}^\mathrm{*,ref}$ values were assumed to be the  $T_{\mathrm{eff}}^\mathrm{*}$ for ZAMS stars provided by  \citet{DMetal98}.
 \item $Q(\mathrm{He}^0)$  values were derived from the corresponding $Q(\mathrm{H}^0)$, $\log g$, and $T_{\mathrm{eff}}^\mathrm{*}$ values from \citet{DMetal98} assuming the continuum shape given by ATLAS atmospheres \citep{Kur91}.
\item A 0.15 M$_{\sun}$ star was included to cover the complete range of stellar masses implicit in the IMF. The $T_{\mathrm{eff}}^\mathrm{*,ref}$ is the  $T_{\mathrm{eff}}^\mathrm{*}$ from a 0.15 M$_{\sun}$ star in the tracks by \cite{Gietal00}. The $Q(\mathrm{H}^0)$ and $Q(\mathrm{He}^0)$ values for this star were obtained by linear extrapolation in the $\log m$ - $\log Q(X)$ plane of the points for 2.6 and 2.9 M$_{\sun}$ obtained in (a) and (b) above. The reason for not using the two lowest mass points of \citet{DMetal98} is the poorer agreement of the linear extrapolation in the $\log m - \log T_{\mathrm{eff}}^\mathrm{*}$ plane from those points with the $T_{\mathrm{eff}}^\mathrm{*}$ of a 0.15 M$_{\sun}$ star. This can be seen in Fig. \ref{fig:m-T} where the $m-T_{\mathrm{eff}}^\mathrm{*,ref}$ relation used in this work and in \citet{DMetal98} are shown.
 \end{enumerate}
 \item High mass stars ($m \geq 20 {\mathrm{M}}_{\sun}$):
 \begin{enumerate}
 \item $T_{\mathrm{eff}}^\mathrm{*,ref}$ values were obtained directly from Cloudy's \citep{Fetal98} implementation of CoStar \citep{SdK97} atmospheres, and the associated evolutionary tracks. The Cloudy inputs were the initial mass and the age (in our case 0.05 Ma  for a ZAMS representation) and provided the corresponding atmosphere model and $T_{\mathrm{eff}}^\mathrm{*}$ from a combination of the CoStar grid.
 \item The $Q(\mathrm{He}^0)$ values  were derived from the resulting atmosphere model assuming a continuum level given by the corresponding  $\log Q(\mathrm{H}^0)$ obtained in step 1 above.
 \end{enumerate}
 \end{enumerate}

The resulting  $m-T_{\mathrm{eff}}^\mathrm{*,ref}$ relation and $T_{\mathrm{eff}}^\mathrm{*,ref}$ vs. $Q(\mathrm{He}^0)/Q(\mathrm{H}^0)$ reference scale are shown in Figs. \ref{fig:m-T} and \ref{fig:Qratio-T}, respectively.

We note that the $T_{\mathrm{eff}}^\mathrm{*,ref}$ vs. $Q(\mathrm{He}^0)/Q(\mathrm{H}^0)$ reference scale depends on the stellar atmosphere code \cite[see][ for an illustrative study]{Metal05,SDS08}. Although this choice is critical to the study of the detailed emission-line spectra of individual stars and  stellar clusters, it has a minor impact on the present study. Our goal is to establish a reference scale to allow a self-consistent comparison of the ionizing properties of individual stars and clusters. Any reference scale is valid as long as it provides a non-degenerate relation between the individual stellar mass (and age) and $Q(\mathrm{He}^0)$,  $Q(\mathrm{H}^0)$, and  their ratio (and, hence, $T_{\mathrm{eff}}^\mathrm{*,ref}$). With this is mind, 
we used ATLAS and CoStar model atmosphere grids,partly also because they are directly available in Cloudy and  are the ones 
used by the evolutionary synthesis code considered in our study.

\subsection{Method}

\begin{figure*}[ht]
\includegraphics[width=\textwidth]{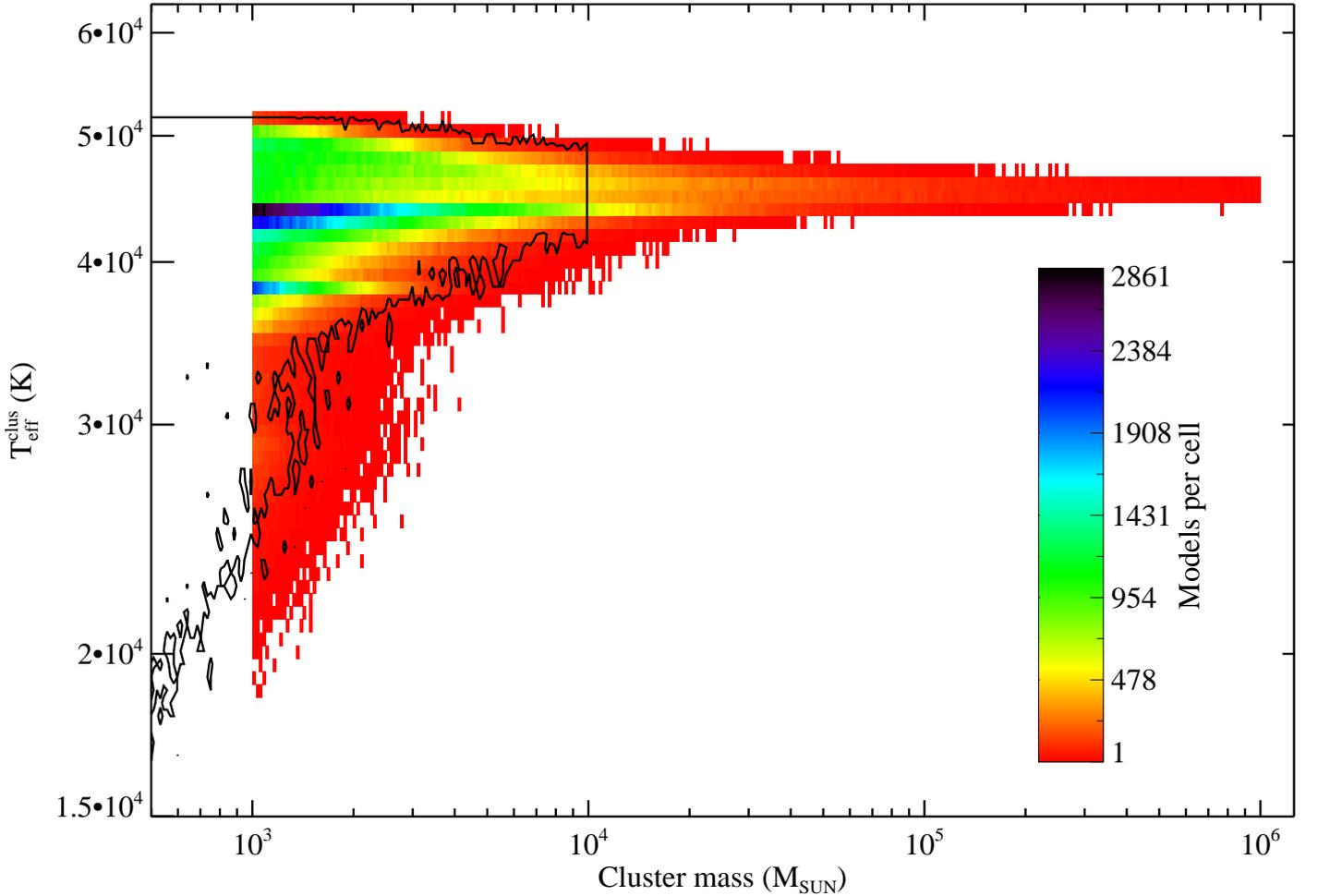}
\caption{The color-coded area shows the $M_\mathrm{clus} - T_{\mathrm{eff}}^\mathrm{clus}$ distribution for the high-mass cluster set. The solid-line contour on the left, reaching  $M_\mathrm{clus}$ lower than $10^3$ M$_{\sun}$, corresponds to the low-mass cluster set (c.f. Fig. \ref{fig:Mcl-Teff_low}).}
\label{fig:Mcl-Teff_high}
\end{figure*}

\begin{figure*}[ht]
\includegraphics[width=\textwidth]{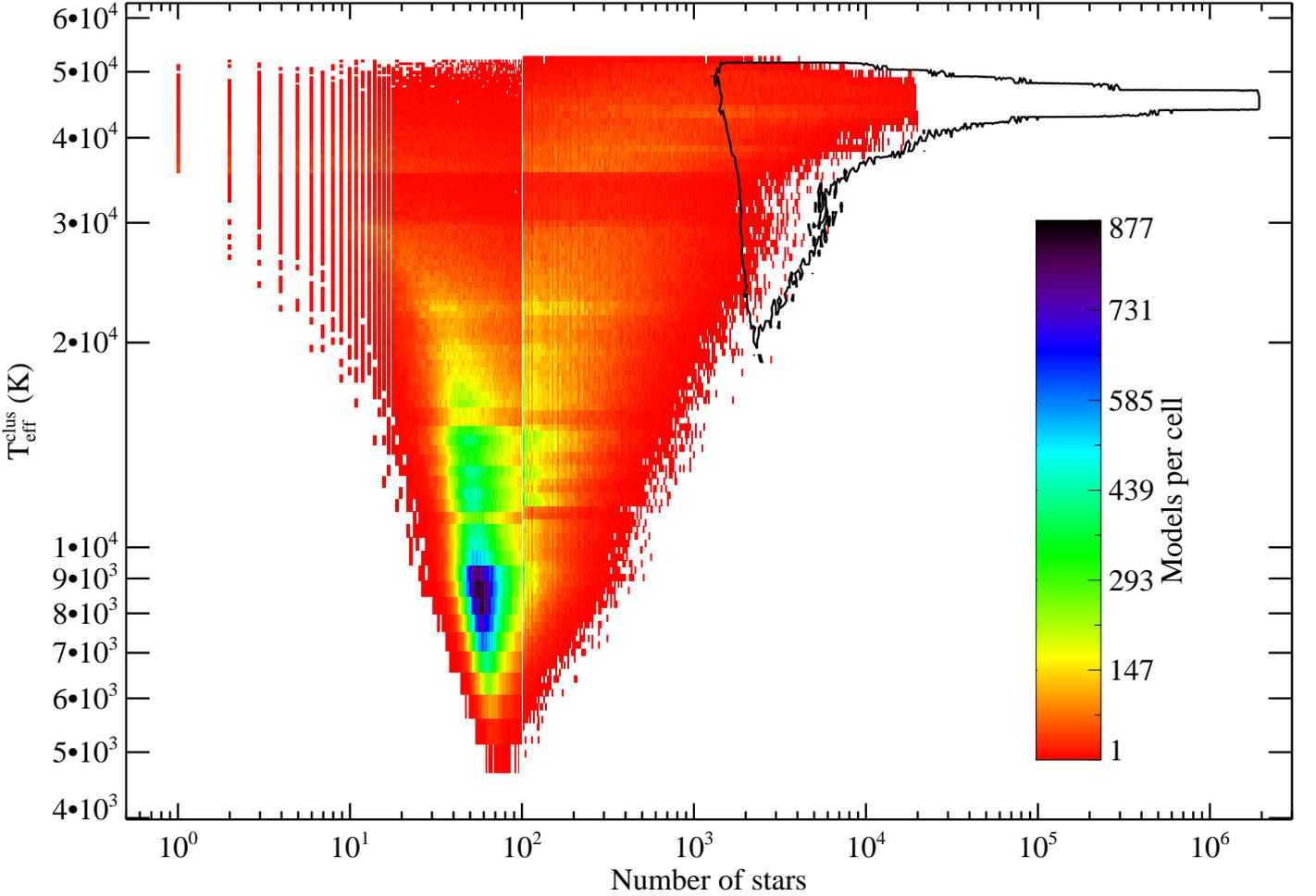}
\caption{The color-coded area represents the ($N_\mathrm{clus}$, $T_{\mathrm{eff}}^\mathrm{clus}$) distribution of the low-mass cluster set. The solid-line contour, extending to  $N_\mathrm{clus}$ larger than $10^5$, shows the distribution of the high-mass cluster set. The vertical discontinuity that appears at $N_\mathrm{clus}=10^2$ is a visual artifact caused by a change from linear binning to logarithmic binning.}
\label{fig:TeffvsN}
\end{figure*}

\begin{figure*}[ht]
\includegraphics[width=\textwidth]{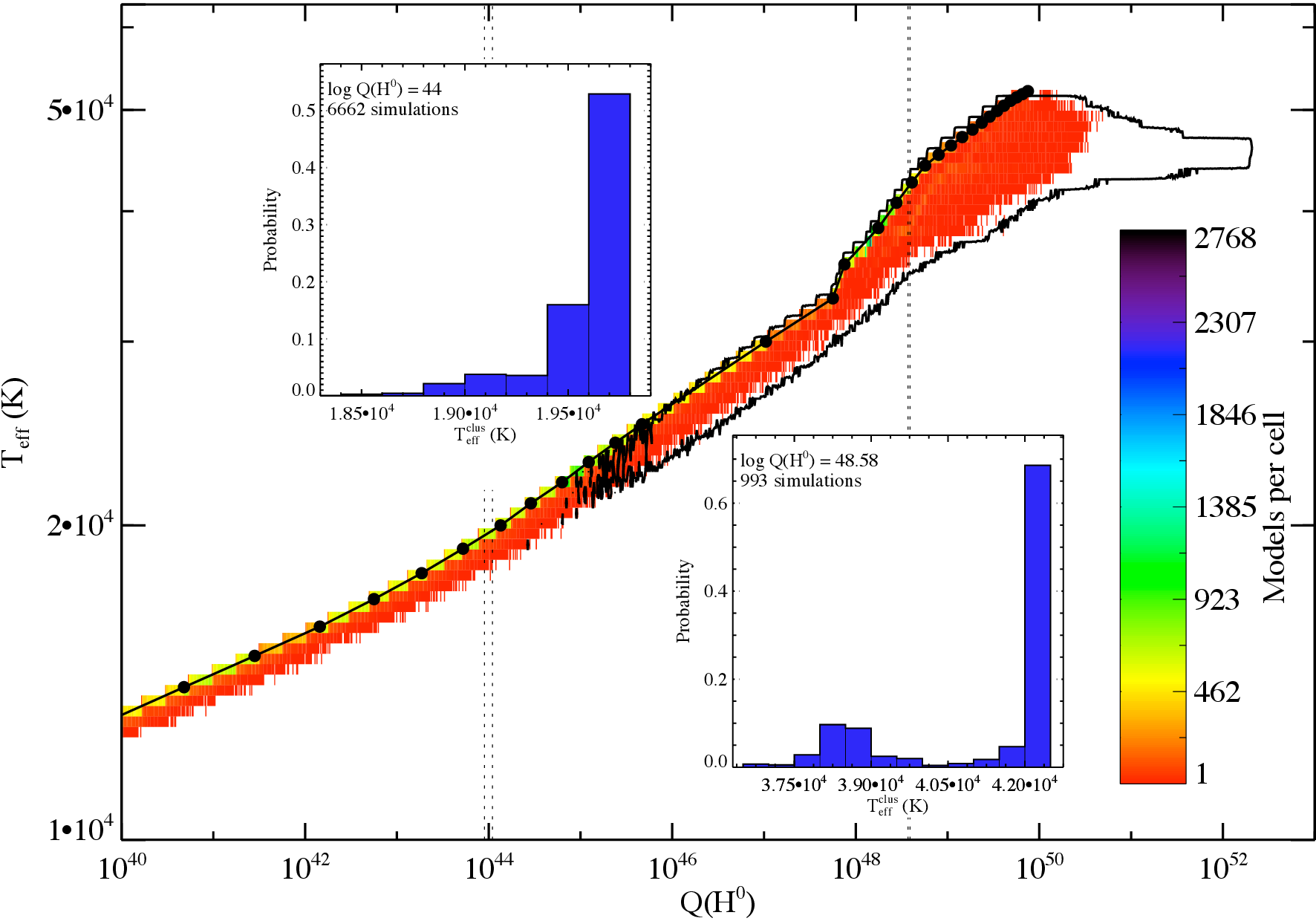}
\caption{Color-coded distribution of the cluster ionizing flux ($Q(\mathrm{H}^0)$)  and $T_{\mathrm{eff}}^\mathrm{clus}$ for the low- and high-mass cluster set. The shaded contour represents the low-mass cluster set, the solid-line contour the high-mass cluster set. A strong correlation between both quantities is seen for the low-mass cluster set case. The same corelation can be seen in the lower part of the distribution of the high mass set, and is lost in the upper part, where the distribution tends to the asymptotic  $T_{\mathrm{eff}}^\mathrm{clus}$.  The insets show the histograms of the $T_{\mathrm{eff}}^\mathrm{clus}$ distribution for two values of $Q(\mathrm{H}^0)$. The range in $Q(\mathrm{H}^0)$ for those histograms are marked with vertical dotted lines.  As a reference, we also plot the $Q(\mathrm{H}^0) - T_{\mathrm{eff}}^\mathrm{*,ref}$ reference scale (line with symbols).} 
\label{fig:QvsTeff}
\end{figure*}

We computed two sets of simulations. The first set was composed of $10^6$ Monte Carlo simulations of clusters with masses from 20 to 10$^4$ M$_{\sun}$ (in the following, the low-mass cluster set). The second set consists of $10^6$ Monte Carlo simulations of clusters with masses from 10$^3$ to 10$^6$ M$_{\sun}$ (high mass cluster set). We note that there is an overlap between the mass range of the two sets of simulations.

Preliminary cluster masses, $M_\mathrm{cl; 0}$, were obtained by means of random sampling of the initial cluster mass function (ICMF) in the cluster mass ranges defined for each set. The assumed ICMF has the form $dN/dM_\mathrm{cl; 0} \propto M_\mathrm{cl; 0}^{-2}$ for all the mass cluster range considered. This representation of the ICMF was proposed by \citet{LL03} for clusters with masses between $\sim$50 and $10^3$ M$_{\sun}$. The application of this ICMF to the complete mass range used here is supported by the studies of \citet{ZF99} and \citet{Hetal03}. 
 
Once a cluster mass was assigned, we used a \cite{Sal55} stellar initial mass function (IMF)   with a lower mass limit $m_\mathrm{low} = 0.15$ M$_{\sun}$ and an upper limit of $m_\mathrm{up} = 100$ M$_{\sun}$, to determine the masses of the individual stars, $m$, in the cluster by performing random sampling of the IMF. The sampling of the IMF continues until the preliminary cluster mass $M_\mathrm{cl; 0}$ is reached. Hence, all simulated cluster have masses $M_\mathrm{clus}$ higher than the preliminary $M_\mathrm{cl; 0}$ and the corresponding ICMF of a given set of Monte Carlo simulations does not follow exactly a power law with exponent $-2$.

In each individual cluster, the $Q(\mathrm{H}^0)$ and $Q(\mathrm{He}^0)$ values of each star of mass $m$ were obtained by linear interpolation of the $\log m$ - $\log Q(X)$ relation obtained previously.  We obtained the cluster's integrated $Q(\mathrm{H}^0)$ and $Q(\mathrm{He}^0)$  by adding the contribution of each individual star, and obtained the cluster's $T_\mathrm{eff}^\mathrm{clus}$ from a linear interpolation in the  $\log Q(\mathrm{He}^0)/Q(\mathrm{H}^0) -  T_\mathrm{eff}^\mathrm{star}$ relation shown in Fig. \ref{fig:Qratio-T}.

We rebin neither the stellar masses nor the cluster masses, so our results correctly map the underlying distribution of physical properties \cite[see][ for different ways of implementing the Monte Carlo simulation]{CL04}.

\subsection{Results}
\label{MCresults}

We now present the resulting multi-dimensional distribution focusing on  $M_\mathrm{clus}$, the number of stars in the cluster ($N_\mathrm{clus}$), $T_{\mathrm{eff}}^\mathrm{clus}$, and $Q(\mathrm{H}^0)$. We compare the results of both sets of simulations using, when possible, the cluster distributions with the reference scale assumed. We also analyse the statistical properties of the distribution of $Q(\mathrm{H}^0)$ and $T_{\mathrm{eff}}^\mathrm{clus}$ as a function of the cluster mass.

\subsubsection{The $M_\mathrm{clus}-T_{\mathrm{eff}}^\mathrm{clus}$ distribution}

The color-coded area in Fig.~\ref{fig:Mcl-Teff_low} shows the bi-parametric $M_\mathrm{clus}-T_{\mathrm{eff}}^\mathrm{clus}$ distribution of the low-mass cluster set. For comparison, the contour of the $M_\mathrm{clus}-T_{\mathrm{eff}}^\mathrm{clus}$ distribution of the high-mass cluster set is overplotted as a black solid line. Figure~\ref{fig:Mcl-Teff_high} shows the same distribution for the high-mass cluster set but with the opposite coding (color-coded for the high-mass cluster set, black solid line for the low-mass cluster set).

The sharp vertical edge in the distributions is due to the minimum cluster mass assumed in the  simulations sets (20 M$_{\sun}$ for the low-mass, and $10^3$ M$_{\sun}$ for the high-mass cluster sets). The solid line with dots on the left-hand side of Fig.~\ref{fig:Mcl-Teff_low} is the $m - T_{\mathrm{eff}}^\mathrm{*,ref}$ reference scale, already plotted in Fig.~\ref{fig:m-T}. 

The $m - T_{\mathrm{eff}}^\mathrm{*,ref}$ reference scale coincides with the low mass envelope of the $M_\mathrm{clus}-T_{\mathrm{eff}}^\mathrm{clus}$ distribution for $M_\mathrm{clus}$ lower than the upper limit of the stellar IMF ($m_\mathrm{up}$). This result is obvious since the $M_\mathrm{clus}-T_{\mathrm{eff}}^\mathrm{clus}$ distribution is limited by the extreme cases of clusters composed by a individual star with mass $m=M_\mathrm{clus}$ (see also Fig. \ref{fig:TeffvsN} below).
The corresponding $T_{\mathrm{eff}}^\mathrm{clus}$ of a cluster composed of a mixture of stars is not so obvious. In general, the resulting $T_{\mathrm{eff}}^\mathrm{clus}$ of a cluster is between the minimum and maximum $T_{\mathrm{eff}}^\mathrm{*,ref}$ of the stars present in the cluster. However, its exact value depends on the relative luminosity of the cluster components. A cluster containing only a 50~M$_{\sun}$ star will have a $T_{\mathrm{eff}}^\mathrm{clus}$ similar to a cluster with a 50~M$_{\sun}$ star in addition to about 20 stars with 5~M$_{\sun}$ (M$_\mathrm{clus}$ = 150 M$_{\sun}$), since the $Q(\mathrm{He}^0)$ and $Q(\mathrm{H}^0)$ produced by such a cluster is dominated by the 50~M$_{\sun}$ star. However, a cluster with a 50~M$_{\sun}$ star will have a different $T_{\mathrm{eff}}^\mathrm{clus}$ to a cluster with both a 50~M$_{\sun}$ star and a 100~M$_{\sun}$ star since both have a non negligible contribution to $Q(\mathrm{He}^0)$ and $Q(\mathrm{H}^0)$ (although this particular combination has a low probability of existing given the IMF). 
The situation in the case of the maximum  $T_{\mathrm{eff}}^\mathrm{clus}$ ($=T_{\mathrm{eff}}^\mathrm{*,ref}$) is illustrated in the inner box of Fig. \ref{fig:Mcl-Teff_low}. The figure shows the locus of individual simulations in the region around the maximum value of the cluster $T_{\mathrm{eff}}^\mathrm{clus}$. The dashed line shows the value of the maximum $T_{\mathrm{eff}}^\mathrm{*,ref}$, equal to  $5.21\times10^{4}\ \mathrm{K}$. As expected, there are no simulations in which $T_{\mathrm{eff}}^\mathrm{clus}$ is higher than this value but, for the assumed ICMF, there is a sizeable number of clusters in the 100 -- $10^3$ M$_{\sun}$ range with such a high effective temperature, i.e. with at least a 100 M$_{\sun}$ star that dominates the ionizing flux.  As the relative contributions of the stellar mixture becomes more homogeneous among clusters, the $T_{\mathrm{eff}}^\mathrm{clus}$ distribution becomes increasingly narrow, and eventually collapses to the asymptotical value $T_{\mathrm{eff}}^\mathrm{clus}$ obtained by synthesis models \cite[see][ for more details]{CL06,CVG09,BdGC08}, which corresponds to $4.54\times10^{4}\ \mathrm{K}$ for the assumed age, IMF, and $m - T_{\mathrm{eff}}^\mathrm{*,ref}$ reference scale.

The $m-T_{\mathrm{eff}}^\mathrm{*,ref}$ reference scale not only defines the low mass envelope of the $M_\mathrm{clus} - T_{\mathrm{eff}}^\mathrm{clus}$ distribution, but also produces multi-modality in the distribution. Steeper slopes in the $m-T_{\mathrm{eff}}^\mathrm{*,ref}$ reference scale correspond to less populated zones in the $M_\mathrm{clus} - T_{\mathrm{eff}}^\mathrm{clus}$ distribution. As an example, the steeper slope in the 17.5 -- 20 M$_{\sun}$ range of the $m-T_{\mathrm{eff}}^\mathrm{*,ref}$ reference scale, corresponding to the $3-3.5\times10^{4}\ \mathrm{K}$ range, is reflected in a minimum in the density of cluster simulations. 

\subsubsection{The $N_\mathrm{clus}-T_{\mathrm{eff}}^\mathrm{clus}$ distribution}

Figure \ref{fig:TeffvsN} shows  the $N_\mathrm{clus} - T_{\mathrm{eff}}^\mathrm{clus}$ distribution for both sets of simulations (only the contour of the simulations covered by the the high mass cluster set is shown). Both set of simulations produce a pyramidal distribution shape.

The left part of this pyramid is produced by the limitation imposed by the minimum cluster mass in both sets.  The clusters that define the lower $N_\mathrm{clus}$  envelope in the low and high mass simulation sets have masses of between 20 and 100 M$_\odot$ (low mass set) and 1000 M$_\odot$ (high mass set). The low $N_\mathrm{clus}$ end of the simulations shows that the clusters with  only a few stars have in fact a high $T_{\mathrm{eff}}^\mathrm{clus}$: they are formed by only a few high mass stars due to the boundary conditions imposed on the simulations.  

The right part of the pyramid is controlled by the asymptotic trend of the clusters with larger $N_\mathrm{clus}$ towards the asymptotic $T_{\mathrm{eff}}^\mathrm{clus}$ value of clusters with infinite number of stars. 

\subsubsection{The $Q(\mathrm{H}^0)-T_{\mathrm{eff}}^\mathrm{clus}$ distribution}

In Fig. \ref{fig:QvsTeff}, we show the $Q(\mathrm{H}^0)-T_{\mathrm{eff}}^\mathrm{clus}$ distribution  
for both sets of simulations, with the same coding as adopted in Fig.~\ref{fig:TeffvsN}. As a reference, we have also plotted the $Q(\mathrm{H}^0) - T_{\mathrm{eff}}^\mathrm{*,ref}$ relation (black solid line with dots), which defines the upper envelope of the distribution in the low-mass set of simulations. The $Q(\mathrm{H}^0) -T_{\mathrm{eff}}^\mathrm{clus}$ relation is very narrow for simulations with $M_\mathrm{clus} \lesssim 10^4$ M$_{\sun}$. Moreover, for the low-mass set given and a fixed $Q(\mathrm{H}^0)$, the most probable $T_{\mathrm{eff}}^\mathrm{clus}$ value is the one corresponding to an individual (reference) star with this given $Q(\mathrm{H}^0)$, i.e., the most massive star compatible with this $Q(\mathrm{H}^0)$ value. This is shown by the small histograms of $T_{\mathrm{eff}}^\mathrm{clus}$ for simulations with $\log Q(\mathrm{H}^0)$ around 44 (as an example of low $Q(\mathrm{H}^0)$).

At  intermediate $\log Q(\mathrm{H}^0)$ values (for example 48.58, the value from synthesis models for a 100 M$_{\sun}$ cluster also shown in the figure)  there appears to be both bimodal distributions and a  larger scatter\footnote{The large area covered by the distribution in this range indicates that there are two relatively narrow distributions centred on different  $T_{\mathrm{eff}}^\mathrm{clus}$ values, rather than a single unimodal distribution with a larger scatter.} in the $Q(\mathrm{H}^0)-T_{\mathrm{eff}}^\mathrm{clus}$ distribution. At such $\log Q(\mathrm{H}^0)$ values, there are clusters containing either an individual dominant star or a mix of stars that mimic clusters with a well sampled IMF but with a lower $m_\mathrm{up}$ value (i.e. high-mass-star deficient clusters).

Finally, when the $Q(\mathrm{H}^0)$ of the cluster is larger than the $Q(\mathrm{H}^0)$ of any individual possible star (in this case, clusters with masses higher than $10^4$ M$_{\sun}$) the cluster  $T_{\mathrm{eff}}^\mathrm{clus}$ distribution slowly converges to the asymptotic  $T_{\mathrm{eff}}^\mathrm{clus}$ value, obtained by synthesis models that do not take into account sampling effects.

The figure illustrates the effects of  sampling  on observable quantities. 
In the low $Q(\mathrm{H}^0)$ regime, 
if we know the $Q(\mathrm{H}^0)$ of a low mass cluster, we can assign a  $T_{\mathrm{eff}}^\mathrm{clus}$ with some confidence, since both relate to similar (dominant) stars, but we are unable to determine the global properties of the cluster ($M_\mathrm{clus}$ or $N_\mathrm{clus}$) with similar confidence. We can, however, determine information  about {\it particular} stars in the cluster (the most luminous ones). This is the case for clusters below the {\it lowest luminosity limit} \citep{CL04}\footnote{The lowest luminosity limit is defined as the luminosity of the brightest individual star that can be present in the cluster for the given evolutionary conditions.}, where extreme sampling effects occur, information about only these particular stars is accessible, not for the cluster as a whole.

Near the {\it lowest luminosity limit}, there are clusters dominated by individual stars (defining a narrow distribution peaking near the assumed  reference scale) and there are clusters whose emission is dominated by an ensemble of stars. This second kind of clusters can be characterised as clusters deficient in high mass stars. In this case, sampling effects can fool us when the asymptotic values obtained by synthesis models are used to obtain {\it global} properties ($M_\mathrm{clus}$, $N_\mathrm{clus}$, or ages). When we modify $m_\mathrm{up}$ in the IMF to mimic the low luminosity clusters, we lose a sizeable number of clusters dominating an individual massive star. If we consider clusters dominated by an individual massive star, we lose a sizeable number of clusters deficient in high mass stars.

In the regime of large $Q(\mathrm{H}^0)$ (values larger than around 10 times the {\it lowest luminosity limit}) both $T_{\mathrm{eff}}^\mathrm{clus}$ and  $Q(\mathrm{H}^0)$ are defined by the proportionality of stellar types provided by the IMF with the correct  $m_\mathrm{up}$, so they provide information about the cluster as a whole and variations in the numbers of particular stars have little impact on the integrated luminosity.

This strong correlation is present only for observed quantities and is lost when the cluster mass is used: the  $Q(\mathrm{H}^0)$ vs. $M_\mathrm{clus}$ distribution (not plotted) shows a similar scatter as the $M_\mathrm{clus}-T_{\mathrm{eff}}^\mathrm{clus}$ distribution shown in Figs.  \ref{fig:Mcl-Teff_low} and \ref{fig:Mcl-Teff_high}. 

\subsubsection{Statistical properties of $Q(\mathrm{H}^0)$}

\begin{figure}[ht]
\includegraphics[width=8.5cm]{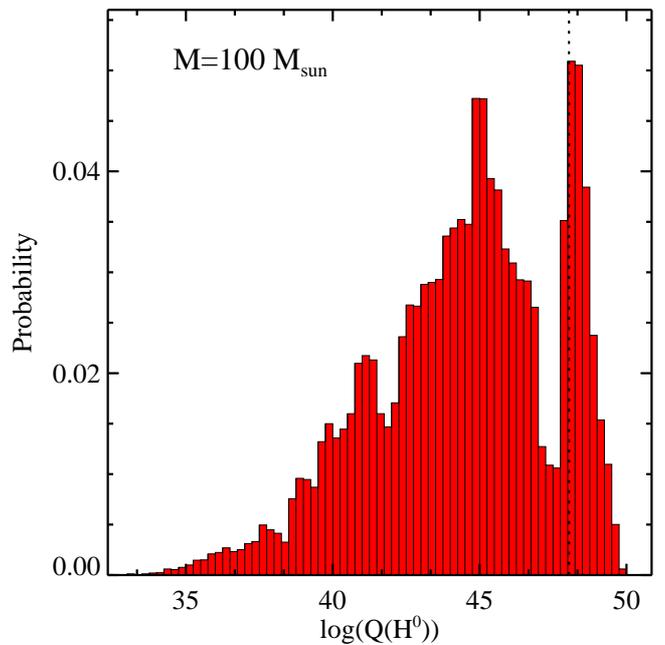}
\caption{Distribution of  $\log Q(\mathrm{H}^0)$ for simulations with $90\,\mathrm{M}_{\sun} \leq M_\mathrm{clus} \leq 110\, \mathrm{M}_{\sun}$ obtained from the low mass cluster set. The probabilities were computed using the simulations in the given mass range (40~502 simulations) and found to have a small dependence on the ICMF. The vertical line shows the mean $\langle Q(\mathrm{H}^0)\rangle$ value ($\log \langle Q(\mathrm{H}^0)\rangle = 48.04$ of the distribution.}
\label{fig:Q-M100}
\end{figure}

Figure \ref{fig:Q-M100} shows the distribution of $\log Q(\mathrm{H}^0)$ in clusters with masses around 100 M$_{\sun}$. The position of the mean value of the distribution ($\langle Q(\mathrm{H}^0)\rangle= 1.1 \times 10^{48}$ ph s$^{-1}, \log \langle Q(\mathrm{H}^0)\rangle = 48.04$) is plotted as a vertical dashed line. We also determined the mean $Q(\mathrm{H}^0)$ for other mass ranges, 
which consistently produces the same number of ionizing photons normalized to mass: $\langle Q(\mathrm{H}^0)\rangle = 1.9 \times 10^{46}$ ph s $^{-1}$ M$_{\sun}^{-1}$ ($\log \langle Q(\mathrm{H}^0)\rangle = 46.27$), 
and yields $\langle Q(\mathrm{H}^0)\rangle= 1.9 \times 10^{48}$ ph s$^{-1}$ for $M_\mathrm{clus}$= 100  M$_{\sun}$.
The discrepancy between this value and the one obtained from the distribution around $M_\mathrm{clus}$=100 M$_{\sun}$ is consistent with low mass clusters being more numerous because of  the ICMF.

Approximately $84\%$ of the clusters have values below the mean, and only $16\%$ above it.  The properties of these overluminous clusters, and therefore their contribution, depend on the upper limit mass of the IMF, $m_\mathrm{up}$, because a change in the upper mass leads to a change in the underlying probability distribution (the very IMF). Therefore, the average value of quantities such as $\langle Q(\mathrm{H}^0)\rangle$ will be not preserved during changes in $m_\mathrm{up}$. In this sense, although there is a strong (statistical) correlation between $M_\mathrm{clus}$ and the star of maximum mass $m_\mathrm{max}$ in the cluster, and an {\it average}~$\langle m_\mathrm{max}\rangle$ can be obtained as a function of $M_\mathrm{clus}$  \citep{WK06}, it is incorrect to use $\langle m_\mathrm{max}(M_\mathrm{clus})\rangle$ as an upper mass limit ($m_\mathrm{up}$) when defining the IMF \cite[whose upper mass limit $m_\mathrm{up}$ appears constant:][]{WK04,MAetal07}. Not only does this change fail to preserve $\langle Q(\mathrm{H}^0)\rangle$): it fails to preserve even $\langle m_\mathrm{max}(M_\mathrm{clus})\rangle$. 

This experiment allows us to illustrate some issues in the modelling of stellar clusters: once an observable is fixed (the total mass of the cluster in this case), the other observables (the number of stars, $Q(\mathrm{H}^0)$, or $T_{\mathrm{eff}}^\mathrm{clus}$ in our case) vary following given probability distributions. In other words, typical relations obtained by synthesis models should be considered valid  {\it as a mean} of the distribution of observed clusters, but not necessarily valid for particular ones. In addition, there are strong correlations between different observables (such as $Q(\mathrm{H}^0)$ and  $T_{\mathrm{eff}}^\mathrm{clus}$), which reduce the scatter in the distributions considerably, especially for undersampled clusters. As a drawback of these situations, the observables do not reflect the cluster properties, but just the properties of the most luminous stars in the cluster.

\subsubsection{Probabilities for given mass ranges}

\begin{figure}[ht]
\includegraphics[width=8.5cm]{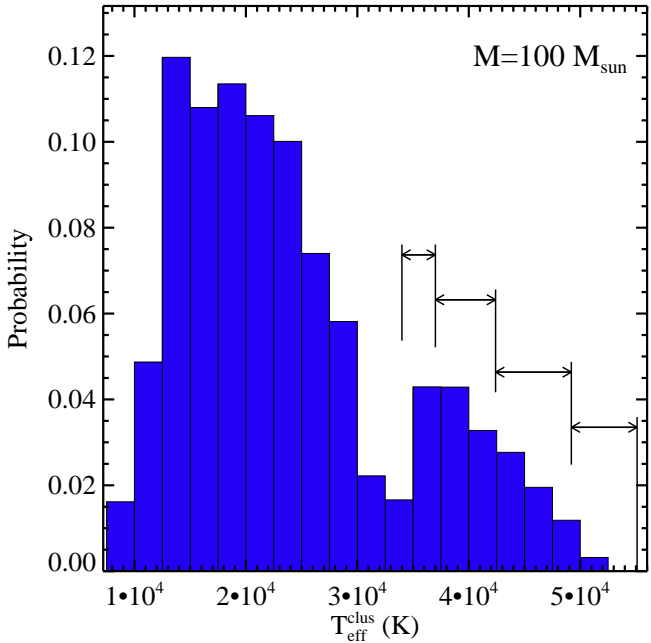}
\caption{Distribution of $T_{\mathrm{eff}}^\mathrm{clus}$ for simulations with $90\,\mathrm{M}_{\sun} \leq M_\mathrm{clus} \leq 110\, \mathrm{M}_{\sun}$ obtained for the low mass cluster set. The probabilities are relative to the number of simulations in the given mass range (40~502 simulations). Vertical lines delimit the regions used for the computation of the probabilities in Table \ref{tab:prob}.}
\label{fig:T-M100}
\end{figure}

\begin{figure}[ht]
\includegraphics[width=8.5cm]{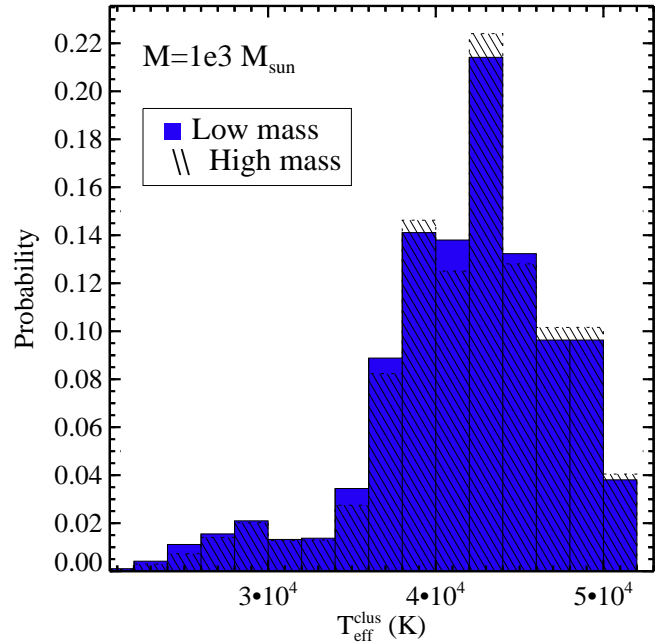}
\caption{Distribution of $T_{\mathrm{eff}}^\mathrm{clus}$ for simulations with $900\,\mathrm{M}_{\sun} \leq M_\mathrm{clus} \leq 1.1 \times 10^3\, \mathrm{M}_{\sun}$ obtained from the low-mass cluster set (4~091 simulations) and $10^3\,\mathrm{M}_{\sun} \leq M_\mathrm{clus} \leq 1.1 \times 10^3\, \mathrm{M}_{\sun}$, obtained from the high-mass cluster set (90~621 simulations).}
\label{fig:T-M1000}
\end{figure}

\begin{figure}[ht]
\includegraphics[width=8.5cm]{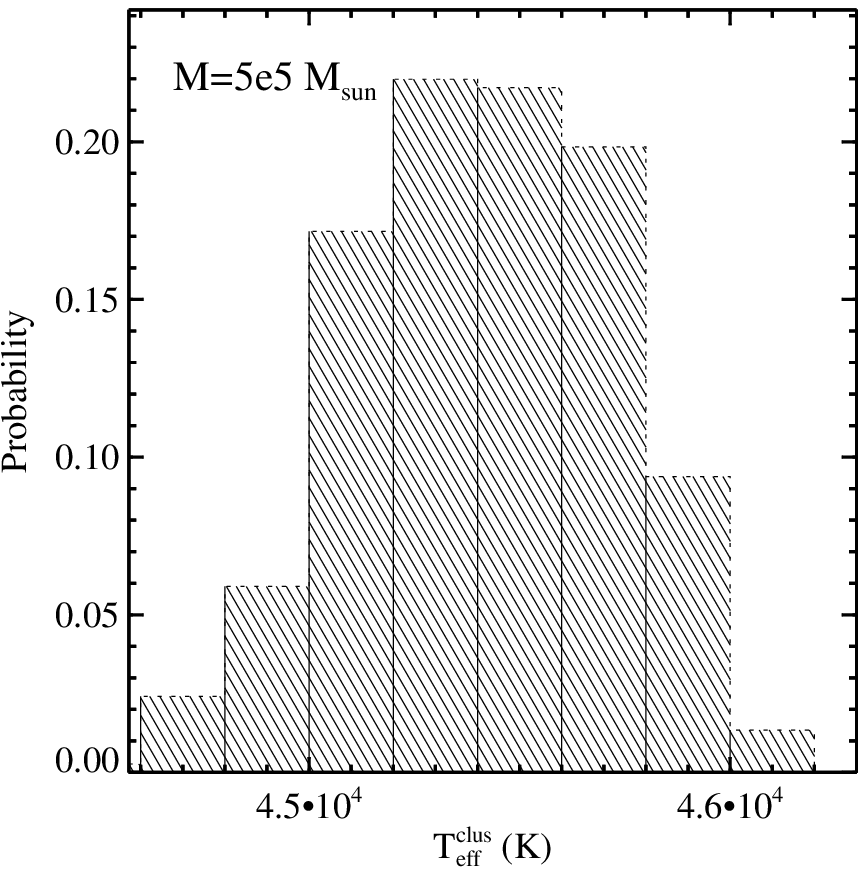}
\caption{Distribution of $T_{\mathrm{eff}}^\mathrm{clus}$ for simulations with $4.5\,\times 10^5 \mathrm{M}_{\sun} \leq M_\mathrm{clus} \leq 5.5\,\times 10^5\, \mathrm{M}_{\sun}$ obtained from the high-mass cluster set (373 simulations).}
\label{fig:T-M5e5}
\end{figure}

We now consider the distribution of the $T_{\mathrm{eff}}^\mathrm{clus}$  for some specific mass ranges.
Figures \ref{fig:T-M100},  \ref{fig:T-M1000}, and \ref{fig:T-M5e5} show three cuts of the distribution for total cluster mass values around 100 M$_{\sun}$, $10^3$  M$_{\sun}$, and $5 \times 10^5$  M$_{\sun}$, respectively. The masses considered are $90\, \mathrm{M}_{\sun} \leq M_\mathrm{clus} \leq 110\, \mathrm{M}_{\sun}$ for Fig. \ref{fig:T-M100}, which cover 40~502 simulations. In the case of Fig. \ref{fig:T-M1000}, which includes simulations from the two sets, we have used $900\,\mathrm{M}_{\sun} \leq M_\mathrm{clus} \leq 1.1\times 10^3\, \mathrm{M}_{\sun}$ in the low mass cluster set with 4~091 simulations, and $10^3\,\mathrm{M}_{\sun} \leq M_\mathrm{clus} \leq 1.1 \times 10^3\, \mathrm{M}_{\sun}$ in the high mass cluster set with 90621 simulations. Finally, the range in  Fig.  \ref{fig:T-M5e5} is $4.5\,\times 10^5 \mathrm{M}_{\sun} \leq M_\mathrm{clus} \leq 5.5\,\times 10^5\, \mathrm{M}_{\sun}$ with 373 simulations. In all cases, a renormalization to the number of simulations in the considered mass range was performed for an easy comparison. 

 \begin{table*}[t]
\centering 
\caption{Probabilities for the simulations with $M\approx 100$ M$_{\sun}$.}
\begin{tabular}{c c c c c}
\hline\hline 
{$T_{\mathrm{eff}}^\mathrm{clus}$ interval (10$^4$ K)} & {\# Simulations} & {Probability (\%)} & {Partial probability (\%)} & Stellar mass (M$_{\sun}$)  \\
\hline
4.92 -- 5.51 & 259 & 0.64 & 3.31 & 100 \\
4.24 -- 4.92 & 2410 & 5.95 & 30.80 & 50 \\
3.70 -- 4.24 & 3503 & 8.65 & 44.77 & 25 \\
3.4 --  3.7 & 1653 & 4.08 & 21.12 & 20 \\
0.2 -- 3.4 & 32677 & 80.68 & - & $\le 20$ \\
\hline
\end{tabular}
\tablefoot{The first column shows the $T_\mathrm{eff}^\mathrm{clus}$ interval in which probabilities are computed; these ranges are also shown in Fig. \ref{fig:T-M100}. The second column gives the number of simulations in each $T_\mathrm{eff}^\mathrm{clus}$ range. The third column shows the probability of each $T_\mathrm{eff}^\mathrm{clus}$ range when the complete set of simulations is considered. The fourth column shows the corresponding probability when only the subset of simulations that can generate an \ion{H}{ii} region is considered. The last column lists the stellar mass corresponding to the middle point of each temperature range.}
\label{tab:prob}
\end{table*}

The $T_{\mathrm{eff}}^\mathrm{clus}$ distribution shown in Fig. \ref{fig:T-M100} span a wide range of values corresponding roughly to the $T_{\mathrm{eff}}^\mathrm{*,ref}$ of stars with masses between 2 and 100 M$_{\sun}$. 
The distribution is bimodal with a mean value of $T_{\mathrm{eff}}^\mathrm{clus}$ around $2\times 10^4$ K, which is clearly biased with respect to the asymptotic value of $4.54 \times 10^4$ K  obtained from the average of the  $T_{\mathrm{eff}}^\mathrm{clus}$ of clusters with $M_\mathrm{clus} > 10^5$ M$_{\sun}$
\cite[see ][for more details of this kind of bias where the observable is obtained from a ratio]{CVG03}.

For $M_\mathrm{clus}$ around $10^3$ M$_{\sun}$, the shape of the distributions is also bimodal, but the mean is more consistent with the asymptotic value. As expected, this bimodality disappears when $M_\mathrm{clus}$ is around $5 \times 10^5$ M$_{\sun}$, where only a small asymmetry remains \cite[cf.][]{CVG03,CL06}.

These histograms allow us to determine the clearest representation of the ionizing continuum by performing a non-exhaustive exploration of the $T_{\mathrm{eff}}^\mathrm{clus}$ distributions. In the following, we describe how focus on the histogram with $M_\mathrm{clus}$ around 100 M$_{\sun}$, which corresponds to the most numerous stellar clusters according to the ICMF proposed by \cite{LL03}. We divided the $T_{\mathrm{eff}}^\mathrm{clus}$ range into five intervals. The middle point of the first four intervals correspond roughly to the $T_{\mathrm{eff}}^\mathrm{*}$ of ZAMS stars with 100, 50, 25, and 20 M$_{\sun}$, the last one including all the clusters with $T_{\mathrm{eff}}^\mathrm{clus}$ lower than $3.4\times10^{4}\ \mathrm{K}$, which we use as an approximate lower limit for the developing of an observable \ion{H}{ii} region. In Table \ref{tab:prob}, the number of simulations and the corresponding probabilities for each interval are shown.
 
Table \ref{tab:prob} shows that most of the zero-aged clusters of 100 M$_{\sun}$ (around 81\%) are unable to generate an \ion{H}{ii} region because they do not have sufficiently massive stars. In the remaining clusters, around 66\% (13\% of the total) have $T_{\mathrm{eff}}^\mathrm{clus}$ below the asymptotic value of $4.54 \times 10^4$ K and only 34\% (6\% of the total)  have a value comparable to or larger than the asymptotic one.

\section{Effects in \ion{H}{ii} regions}

\begin{figure}[ht]
\includegraphics[width=8.5cm]{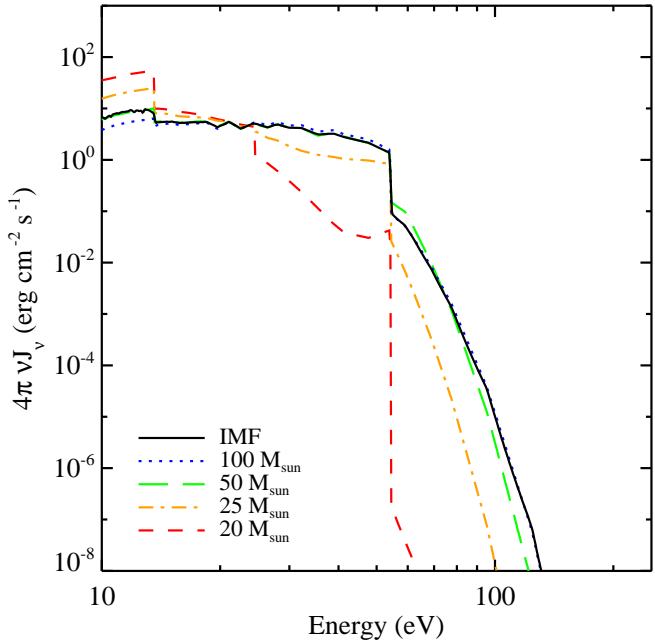}
\caption{Ionizing continua for the scenario of fixed $\log Q(\mathrm{H}^0) = 49.87$. Different lines correspond to different shapes of the ionizing continuum, as explained in the text. Note the small differences among the results of synthesis models (labelled as IMF) and the continuum of a 50 M$_{\sun}$ and 100  M$_{\sun}$ star, consistent with the similarity in $T_\mathrm{eff}^\mathrm{clus}$ between the cluster and star.}
\label{fig:contion}
\end{figure}

The dispersion in $T_{\mathrm{eff}}^\mathrm{clus}$ shown in the previous section implies that there is also a dispersion in the cluster energy distributions. We now show in detail how this dispersion affects the emission-line spectra produced by low mass stellar clusters. We also show how it can render useless the predictions of photoionization grids of clusters modeled as ensembles of stars such as those produced  using synthesis models in a deterministic way, e.g., a univocal value of the  ionizing flux for a given cluster mass.

We focus on  low mass clusters with $M_\mathrm{clus}$ around 100 M$_{\sun}$, which are the most abundant ones. We performed photoionization simulations of this kind of clusters in a simplified way by considering: clusters with 5 ionizing stars ($N_{*,\mathrm{ion}}$) and the ionizing continuum of a 20 M$_{\sun}$ ZAMS  stars, clusters with $N_{*,\mathrm{ion}} = 4$ and the ionizing continuum of a 25 M$_{\sun}$  ZAMS star, clusters with $N_{*,\mathrm{ion}} = 2$ and the ionizing continuum of a  50 M$_{\sun}$ ZAMS star, and clusters with $N_{*,\mathrm{ion}} = 1$ and the ionizing continuum of a  100 M$_{\sun}$ ZAMS star. All ionizing continua have been obtained from CoStar models \citep{SdK97}, which consistent with the $m - T_\mathrm{eff}^\mathrm{*}$ relation described in Sect. \ref{sec:ionMontecarlo}. The summary of this set of photoionization simulations is presented in Table \ref{tab:Mfix}.  We recall that this situation covers only 20\% of all possible cases, since 80\% of clusters of this mass do not produce stars massive enough to develop an H~{\sc ii} region (cf. Table \ref{tab:prob}). 

\begin{table}[tp]
\centering 
\caption{Photoionization simulations for clusters with $M_\mathrm{clus} = 100 \mathrm{M}_{\sun}$.}
\begin{tabular}{c c c c c c}
\hline\hline 
& \multicolumn{4}{c}{Stellar mass} &Complete IMF \\
 &  100 & 50 & 25 & 20 & (0.15-100) \\
 \hline
$N_*$ &   1 & 2 & 4 & 5 & 192.6 \\
$N_{*,\mathrm{ion}}$ & 1 & 2 & 4 & 5 & 0.23 \\
$\log(Q(\mathrm{H}^{0})$ &  49.87 & 49.34 & 48.84 & 48.58 & 48.58  \\
$T_{\mathrm{eff}}^\mathrm{clus} (10^4\:K)$ &  5.01 & 4.53 & 3.81 & 3.53 & 4.60 \\
\hline
\end{tabular}
\label{tab:Mfix}
\end{table}

As a reference value, we also used the ionizing continuum of a solar-metallicity zero-age stellar population synthesis model computed with SED@ \citep{CMH94,Cetal02}\footnote{The SED@ population synthesis models are available at {\tt http://www.laeff.inta.es/users/mcs/SED} with values renormalized to the used 0.15--100 M$_{\sun}$ IMF mass range.}, which uses stellar atmospheres identical to those adopted by ourselves  in  Sect. \ref{sec:ionMontecarlo}. However, SED@ assumes a $m - Q(\mathrm{H}^0)$ relation given by evolutionary tracks, which is different from that used in this work, so that there are some small discrepancies. The most important difference is that  SED@ assumes $m_\mathrm{up} =120$ M$_{\sun}$, hence slightly larger asymptotic values are expected in both $Q(\mathrm{H}^0)$ and $T_{\mathrm{eff}}^\mathrm{clus}$. SED@ provides a $T_{\mathrm{eff}}^\mathrm{clus} = 4.59 \times 10^4$ K, which is 500 K higher than the asymptotic value obtained by the simulations. SED@ also infers that $\log \langle Q(\mathrm{H}^0)\rangle = 3.8 \times 10^{46}$ erg s$^{-1}$ M$_{\sun}^{-1}$, which also differs from the asymptotic value of $1.9 \times 10^{46}$ erg s$^{-1}$ M$_{\sun}^{-1}$ obtained here. However, these discrepancies can be explained by the different value adopted for $m_\mathrm{up}$ in the simulations and SED@,  and do not affect the current discussion.

When synthesis models are used, it is necessary to distinguish between the number of stars in the cluster, $N_*$, and the number of {\it ionizing} stars in the cluster, $N_{*,\mathrm{ion}}$. These numbers can be obtained by simple integrals and do not depend on the synthesis model. The mean number of $N_{*,\mathrm{ion}}$ for a 100 M$_{\sun}$ cluster is 0.23, consistent with the result of Table \ref{tab:prob} where $\sim$ 20\% of the clusters have stars able to develop an H~{\sc ii} region.

We performed additional photoionization simulations for a fixed $Q(\mathrm{H}^0)=49.87$, corresponding to the $Q(\mathrm{H}^0)$ of an individual 100 M$_{\sun}$ star. The summary of these sets is shown in Table \ref{tab:Qfixed}. 

\begin{table}[tp]
\centering 
\caption{Photoionization simulations with $\log Q(\mathrm{H}^0)=49.87$.}
\begin{tabular}{c c c c c c}
\hline\hline 
& \multicolumn{4}{c}{Stellar mass} & Complete IMF  \\
 &  100 & 50 & 25 & 20 & (0.15-100) \\
 \hline
$N_*$ &  1 & 6.76 & 42.66 & 97.72 & 1778.72 \\
$N_{*,\mathrm{ion}}$ & 1 & 6.76 & 42.66 & 97.72 & 2.1 \\
$\log(Q(\mathrm{H}^{0})$ & 49.87 & 49.87& 49.87& 49.87& 49.87\\
$T_{\mathrm{eff}}^\mathrm{clus} (10^4\:K)$ &  5.01 & 4.53 & 3.81 & 3.53 & 4.60 \\
\hline
\end{tabular}
\tablefoot{$\log Q(\mathrm{H}^0)$ equals to $49.87$ corresponds to the $Q(\mathrm{H}^0)$ of an individual 100 M$_{\sun}$ star. }
\label{tab:Qfixed}
\end{table}

As we have already discussed, the tables again show that, in a realistic case, once the cluster mass is fixed, the other properties of the cluster ($N_{*}$, $N_{*,\mathrm{ion}}$, and $Q(\mathrm{H}^0)$ as examples) have distributed values. In the case of $M_\mathrm{clus} = 100$ M$_{\sun}$, there is no solution that would simultaneously satisfy the results in continuum shape, continuum intensity (parametrized as $Q(\mathrm{H}^{0}$)), and cluster mass/number of stars described by the average value obtained from synthesis models. Given that the continuum shape and luminosity control the emission line intensity in a non-linear way \cite[see][ for more details]{VCL09}, using an average ionizing continuum of clusters in function of the cluster mass as input of photoionization codes, as synthesis models provide, gives unrealistic results.

The models with fixed $Q(\mathrm{H}^0)$ allow us to compare the ionizing flux of the different cases within a reference scale. Figure \ref{fig:contion} shows the ionizing continuum for all the ionizing fluxes calculated for the $Q(\mathrm{H}^0)$ fixed case. The shape of the average continua obtained with synthesis models is similar to that of clusters with only 50 M$_{\sun}$ or 100 M$_{\sun}$ stars : the asymptotic cluster $T_\mathrm{eff}^\mathrm{clus}$ is between the $T_\mathrm{eff}^\mathrm{*}$ of a 50  M$_{\sun}$ and that of a 100 M$_{\sun}$ ZAMS  star.
 The figure  also shows the large differences above 24.6 eV in the average continuum obtained from synthesis models and the simulations with 20 and 25 M$_{\sun}$ stars.

The set shown in Table \ref{tab:Qfixed} shows us another interesting conclusion. Synthesis models require about 1800 stars (i.e. cluster masses around $10^3$ M$_{\sun}$) to produce $Q(\mathrm{H}^0)$ similar to an individual 100 M$_{\sun}$ star. This allows us to establish a mass limit for the use of synthesis models in stellar clusters: clusters with masses lower than  $10^3$ M$_{\sun}$ are less luminous than the individual stars that the cluster model contains when described as the average obtained by synthesis models
\cite[the {\it lowest luminosity limit}  established by][]{CL04,Cetal03}. As we have shown (c.f. Fig. \ref{fig:QvsTeff} and discussion in the text), the {\it lowest luminosity limit} indeed establishes  only the minimum cluster mass value to avoid bolometric inconsistencies, although important sampling effects are found for masses 10 times higher than the  {\it lowest luminosity limit} value \citep{CL04}.

\subsection{Effects on the emission line spectrum}

\begin{figure*}[ht]
\centering
\includegraphics[width=\textwidth]{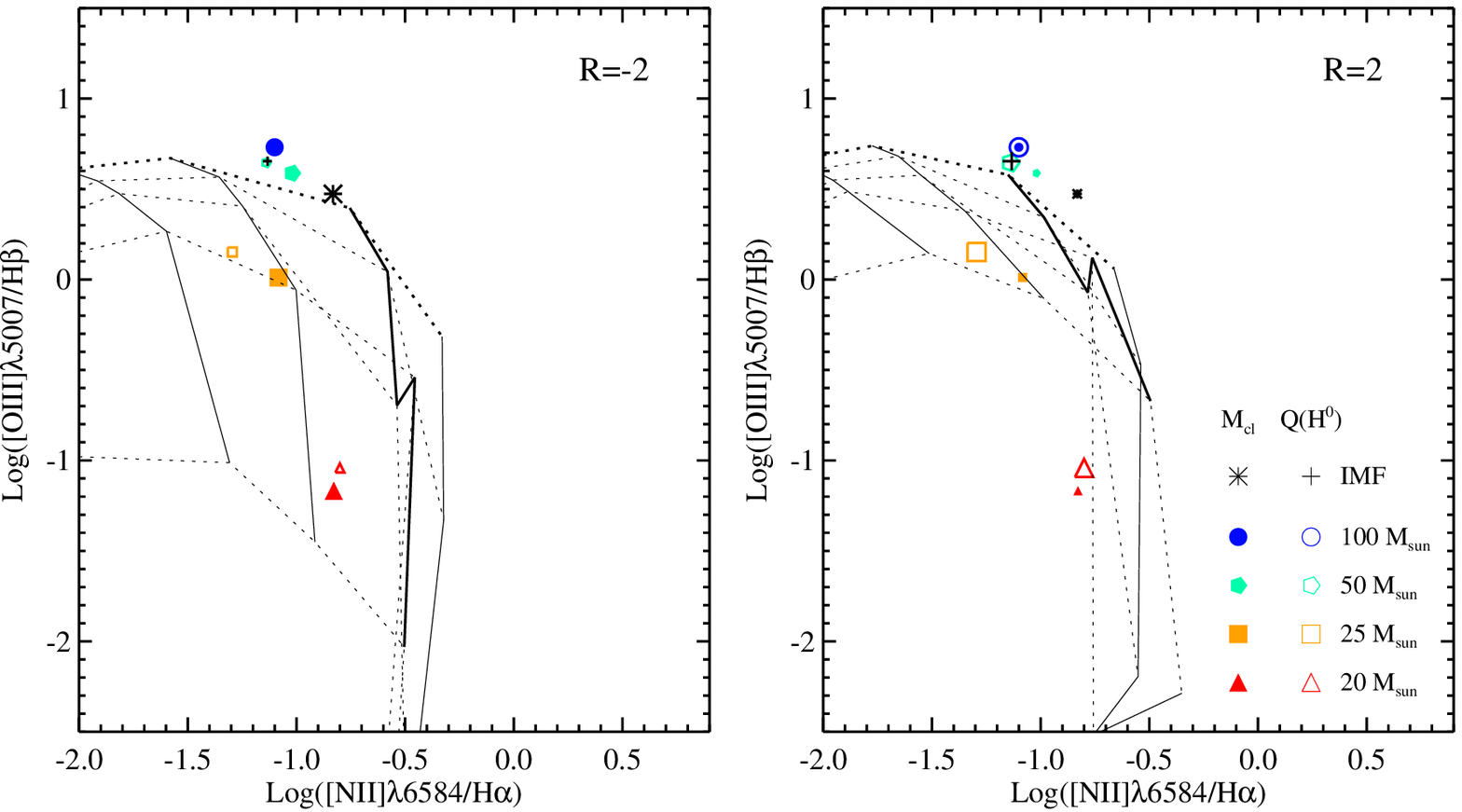}
\caption{$\log $([\ion{O}{iii}] 5007~\AA{} / H$\beta$) vs. $\log $([\ion{N}{ii}] 6584~\AA{}/ H$\alpha$) diagram for the individual-star zero-age solar metallicity simulations. The results are compared with the grid by \cite{Detal06} where dotted lines join points of equal ages, and filled lines join points with equal metallicity. {\it Left:} Grid with logarithm of the ratio of the cluster mass to the ambient presure, ${\cal{R}}$, equals to $-2$, simulations with fixed $M_\mathrm{clus} = 100$ M$_{\sun}$ are plotted as large filled symbols (and models with  fixed $Q(\mathrm{H}^0)$ with small open ones). {\it Right:}  Grid with ${\cal{R}} = 2$, simulations with fixed $\log Q(\mathrm{H}^0) = 49.87$ are plotted as large open symbols (and models with  fixed $M_\mathrm{clus}$ with filled open ones).}
\label{fig:NIIvsOIII}
\end{figure*}

We now study the effects of using an ionizing continuum produced by a cluster consisting of individual stars instead of the average spectrum obtained by synthesis models. We inputted the different continua  into the photoionization code Cloudy \cite[][ version C08]{Fetal98}.  In addition, we assumed a rather simple description of the gaseous component:  an expanding radiation-limited nebula with a 1 pc inner radio, density equal to 100 cm$^{-3}$, and the same solar metallicity mixture as in Table 1 from \cite{Detal06}, although without taking into account any depletion of metals. 

To analyse easily the results obtained, we used diagnostic diagrams and compared with the grids by \cite{Detal06}. These grids were parametrized as a function of a parameter $\cal R$, which is the logarithm of the ratio of the  cluster mass to the ambient pressure. For comparison,  we chose the $\cal R$ that most closely reproduces the position of our Cloudy simulations using the average ionizing continuum obtained by solar metallicity zero-age synthesis models (${\cal{R}} = -2$ for models with fixed $M_\mathrm{clus}$, and ${\cal{R}} = 2$ for models with fixed $Q(\mathrm{H}^0)$ ).

Figure \ref{fig:NIIvsOIII} shows the classical diagram $\log([\mbox{\ion{O}{iii}}] \lambda\,5007 / \mbox{H}\beta)$ \textsl{versus} $\log([\mbox{\ion{N}{ii}}] \lambda\,6584 / \mbox{H}\alpha)$  \cite[][]{BPT} for the two  $\cal R$ values considered. The figure shows large discrepancies when the predictions of the grids are compared with the results of clusters formed with particular stars. In particular, the models with ionizing continua described by 100 and 50 M$_{\sun}$ stars and fixed $M_\mathrm{clus}$ are outside the grid coverage in a region that would correspond to ages younger than those described in the grid (cf. Fig \ref{fig:NIIvsOIII} left). The situation is better for clusters with fixed $Q(H)$ (cf. Fig \ref{fig:NIIvsOIII} right) where the results of synthesis models (labeled as IMF) coincide with the results of clusters with only 50 M$_{\sun}$ stars. In both cases, clusters with ionizing continuum described by 25 and 20 M$_{\sun}$ stars are in the region covered by the grid, although, if the grid is used to estimate ages/metallicities, they are confused with old clusters with a lower metallicity than the one they actually have. The error in the age estimation can be easily explained: the average ionizing continuum produced by synthesis models has a large contribution from the stars at the main-sequence turn-off (the hotter ones in the cluster). As time evolves, the stars at the turn-off become less massive and cooler, so the average cluster ionizing continuum is more accurately described by the 25 and 20 M$_{\sun}$ stars. For clusters affected by sampling effects, the cluster is not actually older, but just deficient in massive stars. This effect would also be reproduced by artificially imposing a lower limit of the $m_\mathrm{up}$ of about 20 M$_{\sun}$: in these cases, the integrated light of the cluster will not vary between zero age and the age when stars with  $m_\mathrm{up}$ begin to leave the main-sequence, and a zero age is compatible with the model result. However, although it would seem to be a good solution for some low-mass clusters (e.g., of 100 M$_{\sun}$), this variation in $m_\mathrm{up}$ is not valid in general (7\% of the clusters have at least a star more massive than 50 M$_{\sun}$, and 80\% of clusters have a maximum $m$ in the particular realization of the IMF that has no stars able to produce an H~{\sc ii} region).

\begin{figure}[ht]
\centering
\includegraphics[width=8.5cm]{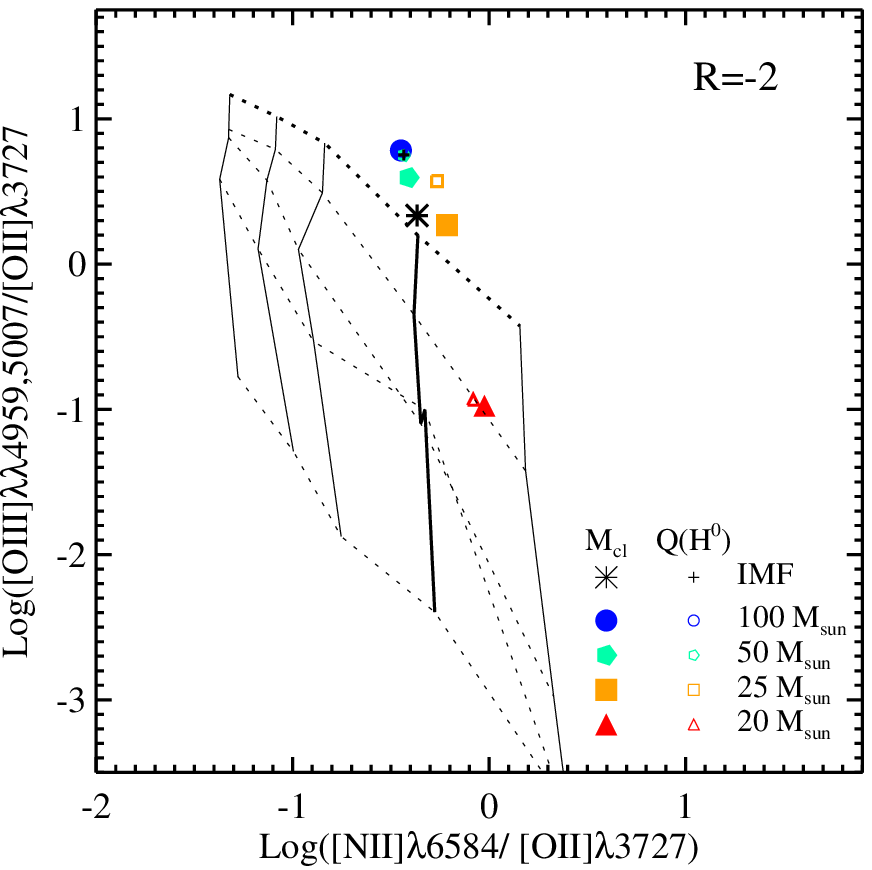}
\caption{Log([\ion{O}{iii}] 5007~\AA{}//[\ion{O}{ii}] 3727, 3729~\AA{}) versus Log([\ion{N}{ii}] 6584~\AA{}/[\ion{O}{ii}] 3727, 3729~\AA{}) for all the simulations. The figure only shows the grid for ${\cal R} = -2$, which corresponds to the cases with fixed $M_\mathrm{clus}$. Lines and symbols as in Fig. \ref{fig:NIIvsOIII}, left. Note that the model of 100 M$_{\sun}$ for fixed $M_\mathrm{clus}$ and the cases of synthesis models (labeled as IMF), 50 and 100 M$_{\sun}$ for fixed $Q(\mathrm{H}^0)$ are in the same position on the plot.
}
\label{O3vsN2b}
\end{figure}
A similar situation can be seen in  Fig. \ref{O3vsN2b},  which shows $\log([\mbox{\ion{O}{iii}}] \lambda\lambda\, 4959,5007 / [\mbox{\ion{O}{ii}}] \lambda\, 3727)$ vs. $\log([\mbox{\ion{N}{ii}}] \lambda\,6584 / [\mbox{\ion{O}{ii}}] \lambda\, 3727)$. This diagram was firstly proposed by \citet{Detal00} for metallicity and ionization parameter estimation  and was also used by \cite{Detal06} for age estimations.
The figure only shows the grid for ${\cal R} = -2$, which corresponds to the cases with fixed $M_\mathrm{clus}$.  The vertical axis again shows a large dispersion because of the differences in the ionizing continuum of the simulations, which can be confused with age variations. On the other hand, the dispersion on the horizontal axis is very small, indicating that the diagram really depends on the gas metallicity.

Variations in the ${\cal R}$ parameter are expected that  for each of the particular cases to porvide a better result. By the same argument, an even better result would be obtained by performing a customized  analysis for each cluster!, taht would only provide information about particular stars in the cluster, but not about the {\it global} cluster properties, as the grids do (excluding the cases of rough metallicity estimations). In conclusion, grids of photoionization models are not useful enough tools for deriving the  properties of stellar clusters affected by sampling effects.

\section{Discussion}

We have desmostrated the limitation of the use of average values obtained by synthesis models to describe the properties of individual clusters with a direct application to H~{\sc ii} regions. Of course, this limitation of the average is proportionally reduced  as the number of stars in the observed cluster increases. However, the situation for low mass stellar clusters \cite[the most abundant ones, following][]{LL03} is complex and the use of the mean value obtained by synthesis models fails.

The main reason for this failure is that the probability of low mass clusters forming massive stars has values around 0.2 in the case of a 100 M$_{\sun}$ zero age stellar cluster from our simulations. We note that this value must be understood statistically rather than interpreted as a value applicable to individual clusters. The correct interpretation of our findings is that 20\% of 100 M$_{\sun}$ zero age stellar clusters have stars massive enough to develop an H~{\sc ii} region (i.e. they have at least one star more massive than 20 M$_{\sun}$) and 80\% of zero age 100 M$_{\sun}$ stellar clusters do not have massive stars and do not form any H~{\sc ii} region at all.  When the clusters are considered individually, this situation poses a severe problem, but when we observe a galaxy of high enough mass the mean value obtained by synthesis models for such a mass provides a good description of the system. 

However, a common misunderstanding of this result is that an individual cluster has 0.2 stars more massive of  20 M$_{\sun}$. Since the number of stars must be a natural number, it can be concluded erroneously that low mass clusters do not form massive stars \cite[the richness effect proposed by][]{GVD94}, or, equivalently, that the upper mass limit of the IMF, $m_\mathrm{up}$, depends on the cluster mass, excluding the cases of clusters formed from molecular clouds less massive than  $m_\mathrm{up}$ \cite[][ and posterior works]{WK06}. This naive interpretation would imply that the ionizing flux (or, equivalently, the H$\alpha$ luminosity) is not a good proxy of the star formation of galaxies, since an {\it ad hoc} cut-off in the upper mass limits of the IMF and the initial cluster mass function (ICMF) is required \cite[c.f.][]{PAWK09}. In terms of $Q(\mathrm{H}^0)$, most low-mass clusters produce less ionizing flux than the mean obtained by synthesis models, but there are also a few low-mass clusters that produce a larger ionizing continuum than the mean given by synthesis models, and the effect cancels out {\it on average} provided there are enough low mass clusters\footnote{Although it is the case for properties such as $Q(\mathrm{H}^0)$, H$\alpha$,  and stellar luminosities whose average values scale linearly with the cluster mass, it would not be  the case for properties that scale in a non-linear way, such as the intensity of collisional emission lines \citep{VCL09}.}.

There are several observational results that are inconsistent with variations in the IMF upper mass limit, $m_\mathrm{up}$, as a function of the cluster mass and confirm the probabilistic interpretation of the IMF  \cite[e.g.][ as some examples]{Coretal09,Jetal04,MAetal07}.
Even more, even if there were clusters with  masses $M_\mathrm{clus}$ lower than $m_\mathrm{up}$ (which implies a hard physical condition for the maximum stellar mass in the IMF {\it realization}), $m_\mathrm{up}$ would be {\it statistically} preserved if  the cluster system produced {\it by the same molecular cloud} contains clusters with $M_\mathrm{clus}$ higher than $m_\mathrm{up}$ \cite[size-of-sample effect,][]{SM08}. 

With our photoionization simulations, we have illustrated the diversity of spectra that the scatter in $T_{\mathrm{eff}}^\mathrm{clus}$ implies and its implications in diagnostic diagrams. The large scatter shown in the diagnostic diagrams is caused only by the way in which we combine stars to obtain the ionizing continuum, which is compatible with a poor IMF sampling. They are not due to metallicity or age effects since the same abundances and ages have been used for all the simulations. When a comparison with theoretical grids of \ion{H}{ii} regions based on the average ionizing continuum produced by synthesis models is performed,  our theoretical clusters may be found to span a wide range in metallicity and/or age.

\subsection{The best ionizing continuum for low mass cluster modelling}

Our results have shown that obtaining the {\it global} properties of low mass clusters, such as their mass or age, is more difficult than expected, since the emission is dominated by individual stars. This situation turns out to be an advantage if we are interested in obtain a good description of the ionizing continuum of the cluster regardless of its evolutionary properties, since the properties of individual stars are better known than the statistical properties of clusters.

In the particular case of the shape (parametrized as $T_\mathrm{eff}^\mathrm{clus}$) and luminosity (parametrized as $Q(\mathrm{H}^0)$) of the ionizing continum, the average obtained by synthesis models is not found be an optimal representation of low mass clusters. At low masses, the cluster $T_\mathrm{eff}^\mathrm{clus}$ and $Q(\mathrm{H}^0)$ are strongly correlated and follow the $T_\mathrm{eff}^\mathrm{*,ref} - Q(\mathrm{H}^0)$ relation of individual stars. That is, the ionizing continua of low mass clusters ($M_\mathrm{clus}  \lesssim 10^4$ M$_{\sun}$), including the case of unresolved ones, are statistically dominated by {\it individual} stars.

{\it As a rule of thumb, we recommend the use of individual stars (with its corresponding stellar $T_\mathrm{eff}^\mathrm{*}$ and  $Q(\mathrm{H}^0)$ values) rather than stellar ensembles (such as those implicit in synthesis models) to model the ionizing flux of low mass clusters ($M_\mathrm{clus}  \lesssim 10^4$ M$_{\sun}$).} However, this ionizing flux does not provide information about the global properties of the cluster (especially the cluster mass or age).
This rule can also be restated in the following terms: {\it ionizing low mass clusters (including unresolved ones) are more suitable for massive stellar atmosphere studies (ionizing spectra studies) than for evolutionary ones.} 

\begin{figure}[ht]
\includegraphics[width=8.5cm]{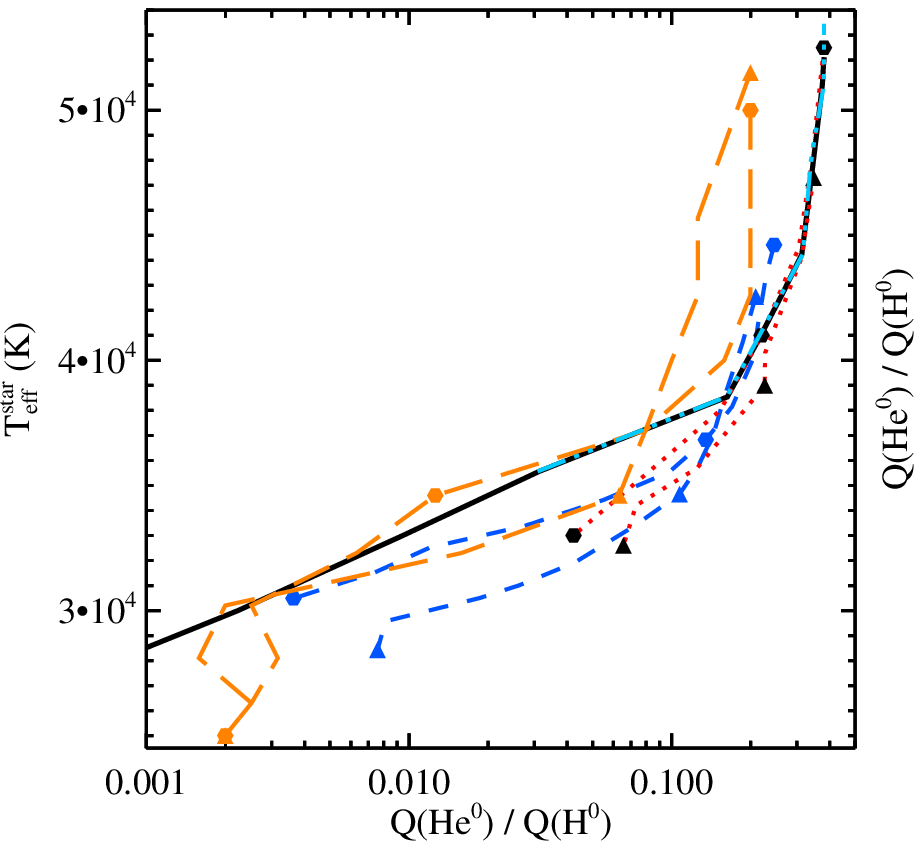}
\caption{ $T_{\mathrm{eff}}^\mathrm{*}$ vs. $Q(\mathrm{He}^0)/Q(\mathrm{H}^0)$  for individual stars assumed in this work based on non-LTE blanketed CoStar models \citep{SdK97} (solid line), \cite{M-HK91} relations based on \cite{Mih72} non-LTE unblanketed models (red dotted line) for stars with luminosity class V and I (hexagons and triangles respectively),  \cite{Metal05} relations based on CMFGEN \citep{HM98} non-LTE spherical extended line-blanketed models (blue short-dashed line) for V and I stars, and \cite{SNC02} relations based in WMbasic \citep{PHL01} computations (orange long-dashed line). }
\label{fig:atmod}
\end{figure}

We note that the relation that links $T_\mathrm{eff}^\mathrm{clus}$ to $Q(\mathrm{He}^{0})/Q(\mathrm{H}^{0})$ has been derived by assuming ZAMS models and a specific set of atmosphere models. A different choice would lead to a different reference scale: generally speaking, any reference scale is atmosphere-model dependent \cite[see][ as examples]{Moretal04,Metal05,SDS08}.
However, any reference scale would be qualitatively similar and lead to similar results that depend on only two results: (a) the IMF in low mass clusters is poorly sampled (so the number of ionizing stars is intrinsically low, and it is just one in low mass ionizing clusters) and (b) the relation between the stellar mass and its ionizing properties is non-degenerate (in such a way that sampling effects in the mass translate directly into the ionizing properties). As long as (b) holds for a particular calibration, the outcome will be the same. This is generally true, as illustrated by Fig. \ref{fig:atmod}.

Although our reference scale has been compiled with particular stellar models, we recall that $T_\mathrm{eff}^\mathrm{clus}$ is not the $T_\mathrm{eff}^\mathrm{*}$ of any particular star, but rather a conventional measure of the cluster Zanstra-like temperature \citep{Zan27}.
That is, the scale can be used as a conventional one and applied to evolved clusters, even if in these the most luminous stars are not ZAMS stars and the stars they contain do not necessarily fit into the reference scale (such as Wolf-Rayet ones). But even if we wish to account for evolution, Fig. \ref{fig:atmod} shows that explicit consideration of different luminosity classes would only add some dispersion to the scale. In sum, our scale is conventional but not arbitrary.

In practice, there may be additional difficulties to consider, such as the presence of more than 
one ionizing star in the unresolved cluster. However, even in these cases the correct approach is not using the mean value from synthesis models, but rather add the contribution of several ionizing stars, in such a way that the $T_\mathrm{eff}^{*}$ and $Q(\mathrm{H}^0)$ of the mix reproduces the observed values \cite[e.g. ][]{Jetal04}.

\section{Conclusions}

We have studied the distribution of stellar cluster properties for a given age and metallicity and the influence that the sampling effects of the IMF may have on it. We have also explored the implications for \ion{H}{ii} region spectra and the estimation of cluster and \ion{H}{ii} region properties obtained with grids and diagnostic diagrams.

For this task, we have performed $2 \times 10^6$ Monte Carlo simulations of zero age clusters in the cluster mass range 20 -- 10$^6$ M$_{\sun}$ to estimate the distribution of the cluster properties. Focusing on clusters with masses around 100  M$_{\sun}$ (the most numerous ones), we conclude that only  20\% of these clusters can generate an \ion{H}{ii} region. We also show that the shape of the average spectra produced by synthesis models only represents 8\% of the possibles cases (or 33\% of clusters that can develop an H~{\sc ii} region). However, those clusters have a higher ionizing flux luminosity than the one produced by the average value obtained by synthesis models, so the average values obtained from synthesis models is correct statistically when all possible cases are considered, but not for individual clusters.

We have shown that variations in the upper mass limit of the IMF with the cluster mass mimics
strong sampling effects for some {\it (but not all)} stellar clusters. However, a correct representation of low mass star clusters is not compatible with an {\it ad hoc} dependence of the upper mass limit of the IMF on the cluster mass since it does not cover all possible cases.

We have also found that there is a strong correlation between the cluster ionizing flux (parametrized as $Q(\mathrm{H}^0)$)  and the shape of the ionizing continuum produced by the cluster. This allows us to  describe the ionizing flux of a low mass cluster ($M_\mathrm{clus} \lesssim 10^4$ M$_{\sun}$) as that produced by a {\it individual} star, instead of using an ensemble of stars. 
Thus H~{\sc ii} regions with ionizing stellar cluster masses lower than or about $10^4$ M$_{\sun}$ are suitable for {\it direct} studies of {\it individual} hot-star atmosphere studies even when individual stars are not fully resolved in the cluster.

Using photoionization models, we have shown that it is not useful to use grids of photoionization models to obtain {\it global} properties of low mass clusters, since they produce erroneous results. 

Finally, we have illustrated how IMF sampling effects work and the reasons why the modelling 
by using individual stars of H~{\sc ii}  regions produced by low mass clusters is more correct than the use of the average parameters obtained by synthesis models such as the cluster $T_\mathrm{eff}^\mathrm{clus}$: sampling effects are a trade-off between global properties of the cluster and particular properties of the stars. When the observables are determined by only a few  stars, global properties are difficult to obtain, and when the observables are defined by the ensemble as a whole, the information about particular stars are diluted by the ensemble.

\begin{acknowledgements}
We thank the referee S. Sim\'on-D{\'\i}az for his general comments, and especially his comments about the stellar calibrations vs the adopted reference scale. We also thank J.M. V{\'\i}lchez for comments about the possible impact of this work on hot-star atmosphere modelling. Finally, we especially thank J.M. Mas-Hesse for recalling us the similarities between Mihalas NLTE unblanketed models and current atmosphere models developments. This work has been supported by the Spanish {\it Programa Nacional de  
Astronom\'\i a y Astrof\'\i sica} through FEDER funding of the project  
AYA2004-02703 and AYA2007-64712.
\end{acknowledgements}

\bibliographystyle{aa} 

\end{document}